# Formation of bulk ferromagnetic nanostructured $Fe_{40}Ni_{40}P_{14}B_6$ alloys by metastable liquid spinodal decomposition

Qiang Li [a]

Department of Physics, The Chinese University of Hong Kong

Shatin, N.T., Hong Kong, P.R. China

[a] School of Physics Science and Technology, Xinjiang University,

Urumqi, Xinjiang 830046, P.R. China

(Tel: +86-991-2109621; Fax: +86-991-8582405; E-mail address: qli@xju.edu.cn)

**Abstract**

Nanostructured $Fe_{40}Ni_{40}P_{14}B_6$ alloys ingots of diameter 3~5 mm could be synthesised by a metastable liquid state spinodal decomposition method. The molten Fe40Ni40P14B6 alloy was purified by means of the fluxing technique and thus a large undercooling could be achieved. For undercooling $\Delta T > 260$ K, the microstructure of the undercooled specimen had exhibited liquid state spinodal decomposition in the undercooled liquid state. The microstructure could be described as two intertwining networks with small grains dispersed in them. For undercooling $\Delta T > 290$ K, the overall microstructure of the specimen changed into a granular morphology. The average grain sizes of the small and large grains are ~ 30 nm and ~ 80 nm, respectively. These prepared samples are soft magnets with saturation magnetization Bs ~0.744 T.

# I. Introduction

Over the past two decades, nanocrystalline magnetic alloys whose average grain size d is less than 100 nm have been attracting more and more attention [1]. In 1988 Yoshizawa et al. [2] reported that a soft magnetic nanocrystalline alloy $Fe_{73.5}Si_{13.5}B_9Cu_1Nb_3$ with a homogeneous superfine nanostructure was obtained by crystallizing its corresponding amorphous alloy ribbons that was produced by rapid solidification processing. Based on this route, three main soft magnetic nanocrystalline alloys, Fe-Si-B-Nb-Cu with the trade name FINEMET [2], Fe-M-B-Cu (M=Zr, Nb, Hf, …) with the trade name NANOPERM [3], and (Fe, Co)-M-B-Cu (M=Nb, Hf, or Zr) with the trade name HITPERM [4] alloys have been found. These nanocrystalline magnetic alloys have exhibited more excellent mechanical and magnetic properties such as ultrahigh strength, high hardness, superplasticity and unique magnetic properties in comparison with amorphous and conventional crystalline magnetic alloys [1,5,6,7].

Principally, nanocrystalline alloys can be synthesized by a variety of techniques such as rapid solidification from the liquid state, mechanical alloying, plasma processing, vapor deposition and so on. However, up to now controlled crystallization from the amorphous state seems to be the only available method to synthesize nanocrystalline alloys with attractive soft magnetic properties. The nanocrystalline magnetic alloys under studied originate from the corresponding amorphous alloys produced by rapid solidification processing. Thus they meet the same limitation in size as that of amorphous alloys. Producing industrially useful bulk nanocrystalline materials is still a challenge for materials engineering. In order to produce bulk nanocrystalline soft magnetic alloys, a consolidation processing is in general employed. However, due to the lower metastability of nanocrystalline alloys (~$0.4RT_m$) than that of amorphous alloys (~$0.5RT_m$) [8], this consolidation of nanocrystalline alloys is more difficult than that of amorphous alloys. Thus a full density and well-bonded bulk nanocrystalline alloy is difficult to be produced in the consolidation processing [6]. Porosity and second-phase inclusions frequently degrade the magnetic properties of the final products. Moreover recrystallization and grain growth during high temperature consolidation processes frequently resulted in a wide grain-size distribution in the produced nanocrystalline alloys. Therefore new techniques to produce perfect bulk nanocrystalline magnetic alloys are needed.

Recently, it [9, 10] was found that when a molten eutectic alloy is undercooled to a temperature T that is substantially below its liquidus $T_l$, it undergoes metastable liquid state spinodal decomposition (LSD) to become a system of intertwining liquid networks of wavelength λ. At large undercooling $\Delta T=T_l-T$, λ enters into the nanometer range and the decomposed networks, driven by surface tension, break up into droplets [11]. The system becomes a nanostructure upon crystallization, characterized as bulk, pore-free, and of narrow grain-size distribution [11, 12]. In this work, the same synthesis method was applied to $Fe_{40}Ni_{40}P_{14}B_6$, a known ferromagnetic amorphous alloy with a trade name of METGLAS alloy #2826, to transform it into a bulk nanostructured alloy.

## II. Experimental

$Fe_{40}Ni_{40}P_{14}B_6$ ingots were prepared from elemental Fe (99.98% pure), Ni spheres (99.95% pure), B granules (99% pure), and $Ni_2P$ ingots (99.98% pure). After the right proportions were weighed, they were put into a clean fused silica tube.

Alloying was brought about a rf induction furnace under Ar atmosphere. All the as-prepared raw ingots had a diameter of 3~5 mm.

It was necessary to undercool a $Fe_{40}Ni_{40}P_{14}B_6$ melt to large $\Delta T$ to initiate LSD. A fluxing technique [13] was designated for this task, in which a molten specimen is immersed in an oxide flux at an elevated temperature that serves to remove heterophase impurities from the melt so as to achieve large $\Delta T$.

In the experiment, a raw $Fe_{40}Ni_{40}P_{14}B_6$ ingot and anhydrous $B_2O_3$ flux were put in a clean fused silica tube. Then the whole system was pumped down to $\sim 1 \times 10^{-3}$ Torr by a mechanical pump before being heated up to ~1350 K by a torch. It was held at 1350 K for 4 h. When the high temperature fluxing was over, the whole system was inserted into a furnace, sitting on a thermocouple as shown in Fig. 1. The furnace was preset at a temperature of $T_k$, which was below the $T_l$ of $Fe_{40}Ni_{40}P_{14}B_6$ (=1184 K). As soon as the thermocouple recorded a crystallization event (connected to a computer), the whole system was removed from the furnace and allowed to cool down in air. So $T_k$ is also the kinetic crystallization temperature. We define the kinetic crystallization undercooling as $\Delta T_k = T_l - T_k$. Microstructures of the as-crystallized or undercooled specimens were studied by X-ray analysis, scanning (SEM) and transmission (TEM) electron microscopy. Both the SEM and TEM were equipped with EDX to conduct composition analysis. Magnetic properties of the nanostructured alloys were measured by means of a VSM.

The microstructures of an undercooled melt with small $\lambda$ evolve readily with time at high temperatures [11, 12]. To avoid confusion, in the following only those undercooled specimens that had been annealed at $T_k$ in the furnace for a period longer than 10 min but less than 30 min were chosen for discussions.

**III. Results**

An undercooled specimen with $T_k$=953 K ($\Delta T_k$ = 231 K) comprises two types of grains, classified according to their grain sizes d and compositions (Fig. 2). Most often, the smaller grains exist as inclusions inside the larger ones. The compositions of the small and large grains studied by TEM method and x-ray analysis (details next paragraph) are, respectively, (Fe,Ni) and $(Fe,Ni)_3(P,B)$.

The microstructure of an undercooled specimen with $T_k$ = 924 K ($\Delta T_k$ = 260 K) is shown in Fig. 3a. The constituent grains can again be classified, according to their sizes as well as compositions, into two groups. The smaller ones, with an average grain size smaller than 100 nm and isolated from each other, distribute themselves randomly throughout the entire specimen but preferentially at the grain boundaries of the larger grains. The larger grains with wavy boundaries and average grain size ~200 nm form the background. They appear with different degree of brightness for there is a slight change in their grain orientations. It is most interesting to observe that when neighbouring grains of similar degree of brightness are joined together, the overall microstructure of the undercooled specimen can be described as two intertwining networks with the smaller grains dispersed in them. To determine the compositions of the constituent grains, it was necessary to employ x-ray analysis because B atoms cannot be measured accurately by means of an EDX. A typical x-ray diffraction pattern is shown in Fig. 3b. Combining the electron diffraction analysis, EDX and x-ray results, the compositions of the smaller and larger grains are, respectively, (Fe,Ni) and (Fe,Ni)$_3$(P,B). These phases are identical to those appeared in a crystallized Fe$_{40}$Ni$_{40}$P$_{14}$B$_6$ glassy ribbon [14].

At $T_k$ = 894 K ($\Delta T_k$ = 290 K), the smaller grains or (Fe,Ni) precipitates are seldom found as inclusions in the larger grains of composition, (Fe,Ni)$_3$(P,B) as shown in Fig. 4. The respective average grain size of the small and large grains are ~ 30 nm and ~80 nm. The overall microstructure of the specimen is granular, different from the network-like morphology shown in Fig. 3a.

The saturation magnetization $B_s$ and coercivity $H_c$ of a specimen with $T_k$ = 894 K ($\Delta T_k$=290 K) were measured by a VSM. $B_s$ was ~80 emu/g. Its density can be determined to be 7.4 kg/m$^3$ by means of immersed water method and thus the saturation magnetization $B_s$ of the specimens can be deduced to be 0.744 T which is less than that of an bulk amorphous Fe$_{40}$Ni$_{40}$P$_{14}$B$_6$ alloy, 0.859 T [15]. And $H_c$ is too small to be resolved by the VSM.

## IV.   Discussions

The quartnary Fe$_{40}$Ni$_{40}$P$_{14}$B$_6$ alloy is a typical eutectic amorphous alloy with TM$_{80}$M$_{20}$ form. On deritrification of a melt-spun ribbon, the resulting phases consist of an austenite Ni-Fe solid solution and a body-centered tetragonal (Ni,Fe)$_3$(P,B) [14]. In a DSC measurement, there is only a single-stage exothermic reaction [15]. These results suggest strongly that Fe$_{40}$Ni$_{40}$P$_{14}$B$_6$ can be treated as a pseudobinary eutectic (Fe,Ni)-(P,B) system. And its equilibrium phase diagram should be similar to that of Ni-P binary eutectic alloy and Fe$_{40}$Ni$_{40}$P$_{14}$B$_6$ is very close to the eutectic composition in the phase diagram. Based on the equilibrium phase diagram of Ni-P alloys, a hypothesis equilibrium phase diagram of pseudobinary (Fe,Ni)-(P,B) alloys is shown in Fig. 5.

If a eutectic melt can avoid crystallization, it can be undercooled and enters into its miscibility gap. So the liquid phase separation can be expected to occur and the homogenous liquid will decompose into a few intertwining liquid networks by nucleation and growth mechanism or by spinodal decomposition mechanism depending on the undercooling of the melt [9, 10]. After liquid phase separation, these phase-separated liquids are still undercooled and thus subsequent crystallization will occur in these phase-separated liquids.

When a system of two or more intertwining liquid networks is still connected and crystallization starts out in one network, the crystal growth front would propagate along that network epitaxially in disregard of what the other liquid networks are doing [16, 17, 18]. After all the liquid networks turn crystalline, it is likely that structural relaxation that serves to relieve stress induced by the crystallization process can be brought about by introducing small angle grain boundaries (e.g., from twisting of dendritic arms) along the branches of the networks. That when viewed under the TEM is a network-morphology made up of grains of similar brightness as shown in Fig. 3a. The (Fe,Ni) precipitates (shown in Fig. 2, 3a and 4) appear for the compositions of the metastable liquid networks just before crystallization are close to, but not exactly equal to $(Fe,Ni)_3(P,B)$.

Comparing Fig. 2 with Fig. 3a, there is a transition in microstructural size, displayed vividly by the phase $(Fe,Ni)_3(P,B)$ as well as a transition in morphology, from granular to network-like. Accordingly [10], the microstructure shown in Fig. 2 was due to a liquid state nucleation and growth reaction. Finally the refined and granular morphology shown in Fig. 4 indicates that just before crystallization the liquid networks had successfully broken up into liquid droplets [10,11].

According to the above analysis, a hypothesis miscibility gap for pseudobinary eutectic (Fe,Ni)-(P,B) alloy has been constructed as shown in Fig. 5. At $T_k$=953 K ( $\Delta T_k$ = 231 K), the undercooled eutectic melt will enter into region II as shown in Fig.5. The homogenous liquid phase will decompose into island-like morphology, consisting of phase-separated liquids, by nucleation and growth mechanism. Subsequent crystallization can freeze the island-like morphology. When $T_k$ < 924 K ( $\Delta T_k$ > 260 K), a melt will enter inside the spinodal line, i.e. region I, as shown in Fig. 5. It will decompose into randomly interconnected liquid sub-networks and the wavelength $\lambda$ of sub-network will decrease with the increase of undercooling $\Delta T_k$. The smaller wavelength $\lambda$ results in the larger interfacial areas. The interfacial energy finally lead to the break up of the liquid networks into droplets when $T_k$ = 894 K ( $\Delta T_k$ = 290 K). Subsequently crystallization will result in a perfect nanostructured alloy as shown in Fig. 4.

**Acknowledgement**

I thank Hong Kong Research Grants Council and Xinjiang University Doctoral Research Start-up Grant (Grant no.: BS050102) for financial support.

**References**

[1]  McHenry ME, Willard MA, Laughlin DE, Progress in Materials Science 1999;44:291;

[2]  Yoshizawa Y, Oguma S, Yamauchi K, J. Appl. Phys. 1998;64:6044;

[3]  Kojima HH, Kamamura Y, Makino A, Inoue A, Matsumoto T, Mater. Sci. Eng. A 1994;179-180:511;

[4]  Iwanabe H, Lu B, McHenry ME, Laughlin DE, J. Appl. Phys. 1999;85:4424;

[5]  Herzer G, Scripta Metallurgica et Materialsa 1995;33:1741;

[6]  Weertman JR, Farkas D, Hemker K, Kung H, Mayo M, Mitra R, et al., MRS Bull. 1999;24(2):44;

[7]  Koch CC, Morris DG, Lu K, Inoue A, MRS Bull. 1999;24(2):54;

[8]  Turnbull D, Metall. Trans. A 1981;12:695;

[9]  Yuen CW, Kui HW, J. Mater. Res. 1998;13:3034;

[10] Lee KL, Kui HW, J. Mater. Res. 1999;14:3653;

[11] Guo WH, Chua LF, Leung CC, Kui HW, J. Mater. Res. 2000;15:1605;

[12] Guo WH, Kui HW, Acta Mater. 2000;48:2117;

[13] Kui HW, Greer AL, Turnbull D, Appl. Phys. Lett. 1984;45:615;

[14] Morris DG, Acta Mater. 1981;29:1213;

[15] Li Q, Materials Letters 2006;60:3113;

[16] Lee KL, Kui HW, J. Mater. Res. 1999;14:3663;

[17] Guo WH, Leung CC, Kui HW, MRS Symp. Proc. 1999;554:211;

[18] Leung CC, Guo WH, Kui HW, Appl. Phys. Lett. 2000;77: 64.

**Figure Captions**

Fig. 1　　The experimental setup schematic.

Fig. 2　　Microstructure of an undercooled specimen with $T_k$ = 953 K ($\Delta T_k$ = 231 K). Most of the smaller grains are enclosed inside the larger grains.

Fig. 3a　Microstructure of an undercooled specimen with $T_k$ = 924 K ($\Delta T_k$ = 260 K). Most of the smaller grains are at the boundaries of the larger grains.

Fig. 3b　X-ray diffraction pattern of an undercooled specimen with $T_k$ = 924 K ($\Delta T_k$ = 260 K). The phases are also indicated in the figure.

Fig. 4　　Microstructure of an undercooled specimen with $T_k$ = 894 K ($\Delta T_k$ = 290 K). It can be described as granular.

Fig. 5　　A hypothesis equilibrium phase diagram and its miscibility gap for pseudobinary eutectic (Fe,Ni)-(P,B) alloy

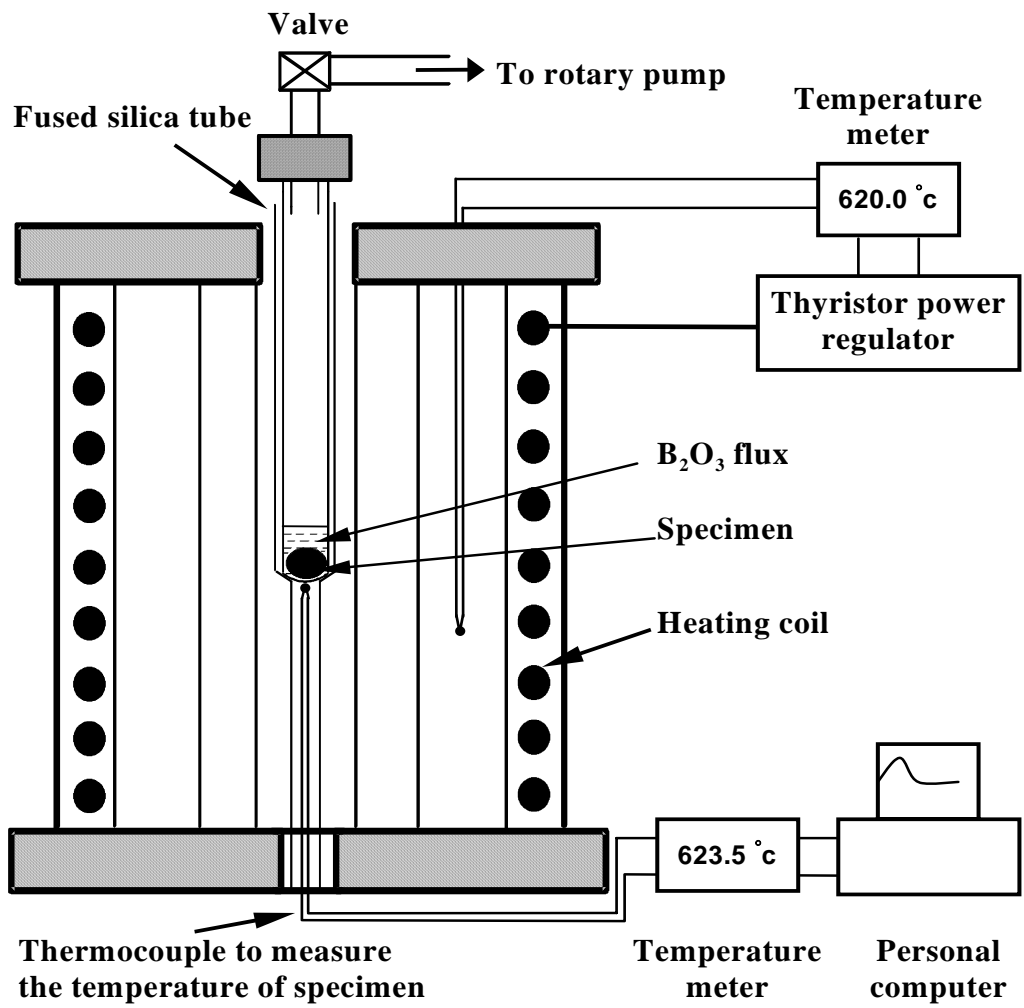

Fig. 1 The experimental setup schematic

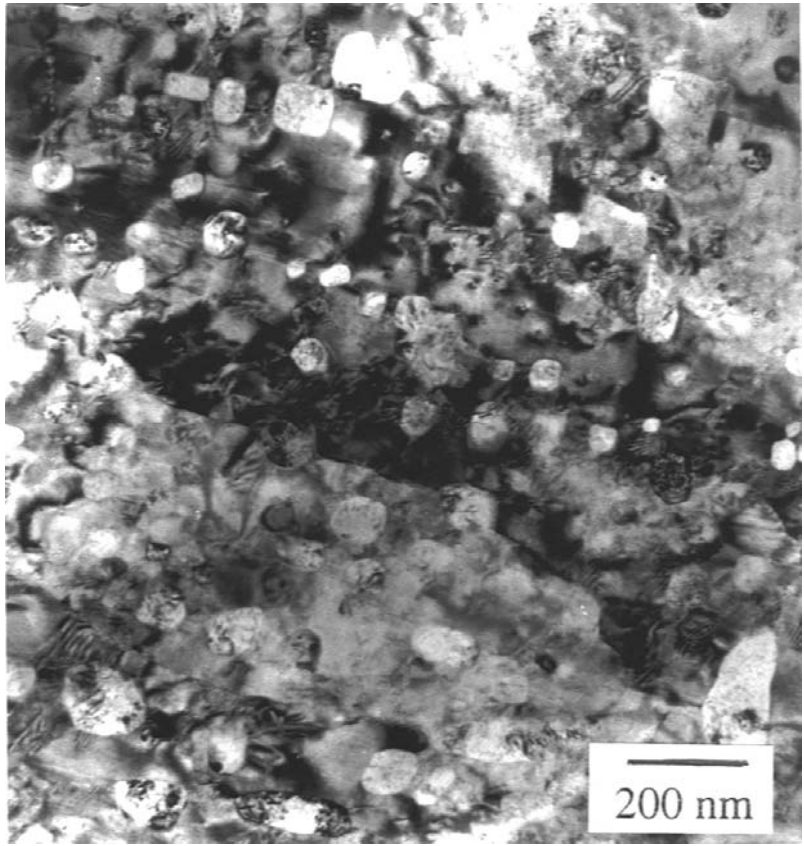

Fig. 2 Microstructure of an undercooled specimen with $T_k = 953$ K ($\Delta T_k = 231$ K). Most of the smaller grains are enclosed inside the larger grains.

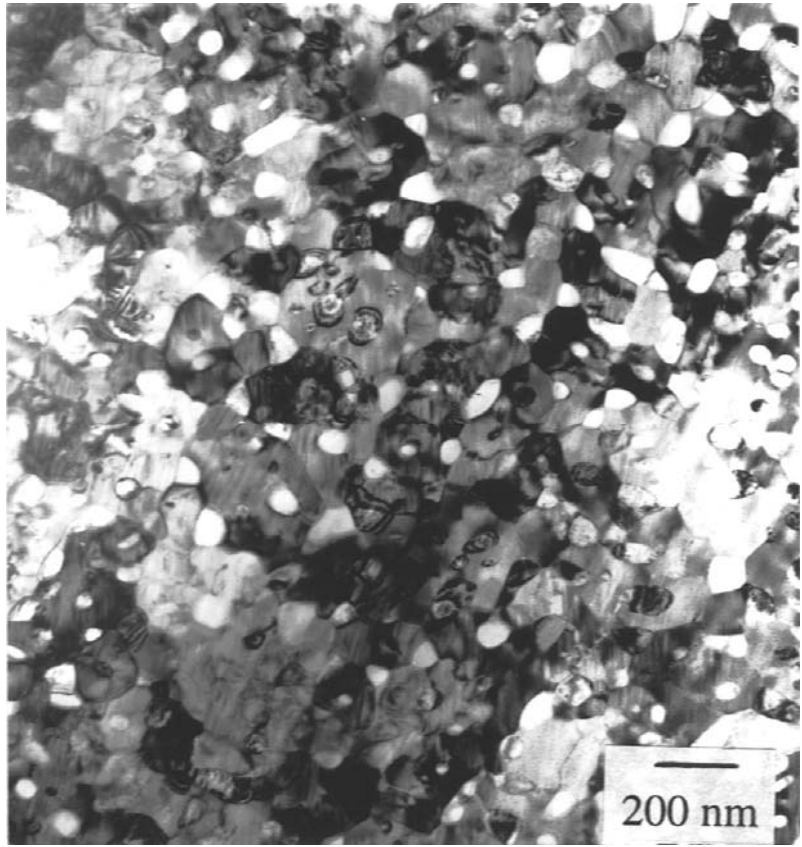

Fig. 3a Microstructure of an undercooled specimen with $T_k = 924$ K ($\Delta T_k = 260$ K). Most of the smaller grains are at the boundaries of the larger grains.

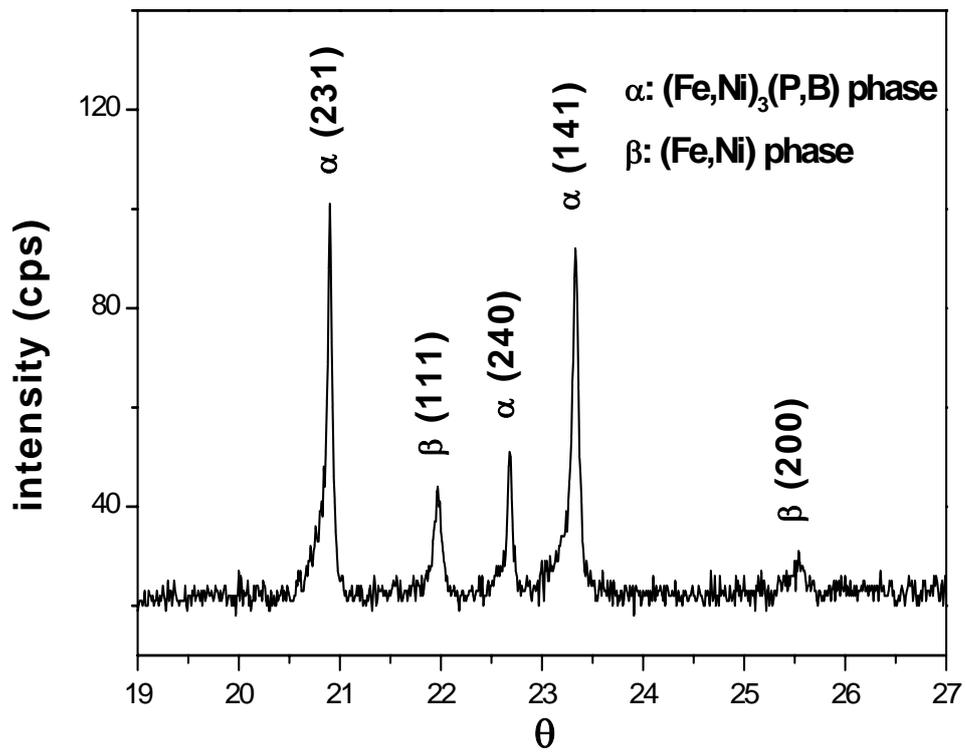

Fig. 3b X-ray diffraction pattern of an undercooled specimen with $T_k = 924$ K ($\Delta T_k = 260$ K). The phases are also indicated in the figure.

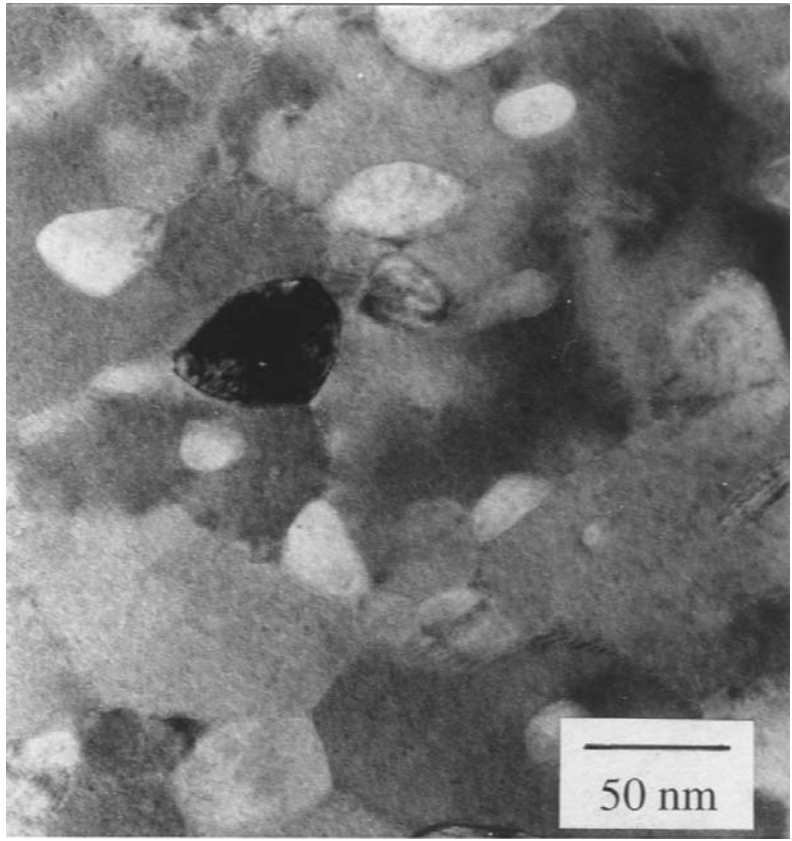

Fig. 4 Microstructure of an undercooled specimen with $T_k$ = 894 K ($\Delta T_k$ = 290 K). It can be described as granular.

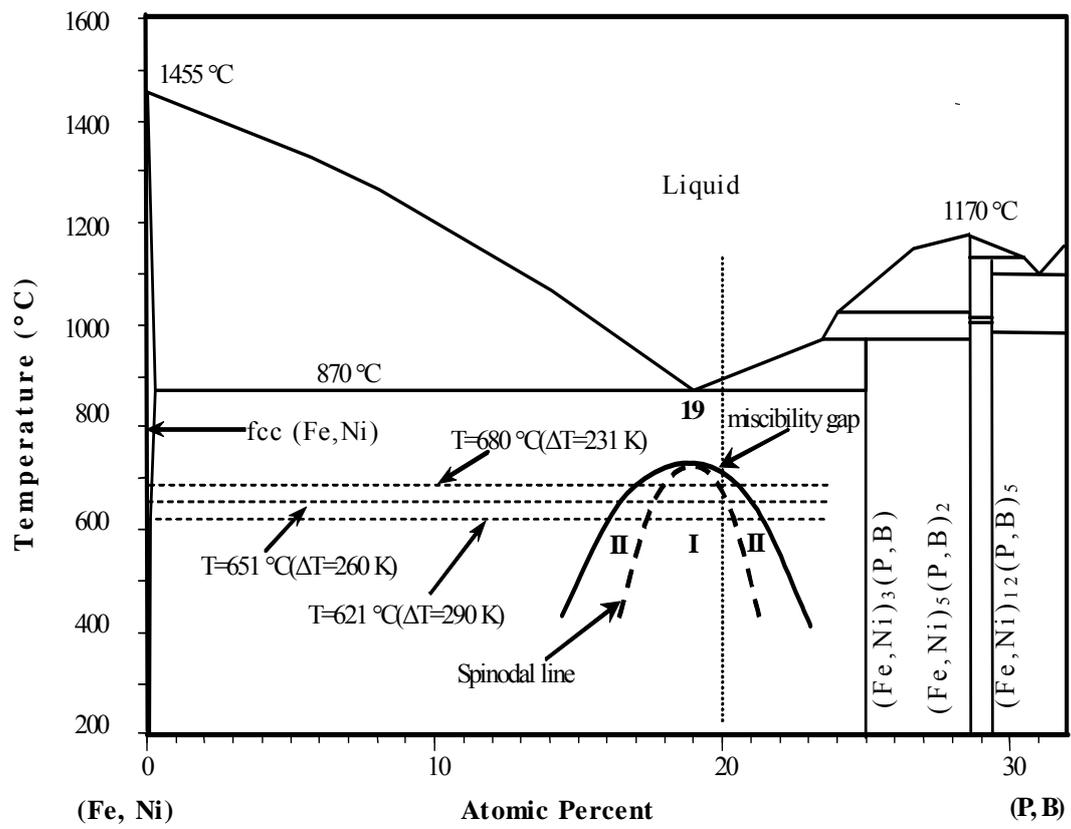

Fig.5 A hypothesis equilibrium phase diagram and its miscibility gap for pseudobinary eutectic (Fe,Ni)-(P,B) alloy